# A Review on the Salt Bridge ASP177-ARG163 (O–N) of Wild-Type Rabbit Prion Protein

Jiapu Zhang[ab*], Feng Wang[a]

[a]Molecular Model Discovery Laboratory, Department of Chemistry & Biotechnology, Faculty of Science, Engineering & Technology, Swinburne University of Technology, Hawthorn Campus, Hawthorn, Victoria 3122, Australia;

[b]Graduate School of Sciences, Information Technology and Engineering & Centre of Informatics and Applied Optimisation, Faculty of Science, The Federation University Australia, Mount Helen Campus, Mount Helen, Ballarat, Victoria 3353, Australia

[*]Correspondence address: Telephones: +61-3-9214 5596, +61-3-5327 6335, +61-423 487 360; E-mails: jiapuzhang@swin.edu.au, j.zhang@federation.edu.au, jiapu_zhang@hotmail.com

**Abstract:**

Prion diseases are invariably fatal and highly infectious neurodegenerative diseases that affect a wide variety of mammalian species such as sheep and goats, cattle, deer, elks, humans and mice etc., but rabbits have a low susceptibility to be infected by prion diseases with respect to other species. The stability of rabbit prion protein is due to its highly ordered β2-α2 loop (PLoS One 5(10) e13273 (2010); Journal of Biological Chemistry 285(41) 31682-31693 (2010)) and a hydrophobic staple helix-capping motif (PNAS 107(46) 19808-19813 (2010); PLoS One 8 (5) e63047 (2013)). The β2-α2 loop and the tail of Helix 3 it interacts with have been a focus in prion protein structure studies. For this loop we found a salt bridge linkage ASP177-ARG163 (O-N) (Journal of Theoretical Biology 342 (7 February 2014) 70-82 (2014)). Some scientists said *on the 2FJ3.pdb NMR file of the rabbit prion protein, the distance of ASP177-ARG163 (O-N) gives the salt bridge of about 10 Å which is nearly null in terms of energy* and *such a salt bridge is not observed in their work.* But, from the 3O79.pdb X-ray file of the rabbit prion protein, we can clearly observe this salt bridge. This article analyses the NMR and X-ray structures and gives an answer to the above question: *the salt bridge presents at pH 6.5 in the X-ray structure is simply gone at pH 4.5 in the NMR structure* is simply due to the different pH values that impact electrostatics at the salt bridge and hence also impact the structures. Moreover, some molecular dynamics simulation results of the X-ray structure are reported in this article to reveal the secrets of the structural stability of rabbit prion protein.

**Keywords:** prion diseases; rabbit prion protein wild-type and mutants; X-ray and NMR structures; molecular dynamics study; loop β2-α2; salt bridge; drug target

**Abbreviations**: $PrP^C$, a soluble normal cellular prion protein; $RaPrP^C$, rabbit $PrP^C$; $PrP^{Sc}$, insoluble abnormally folded infectious prions; BSE, bovine spongiform encephalopathy; TSE, Transmissible Spongiform Encephalopathy; CJD, Creutzfeldt-Jakob Disease; vCJD, variant Creutzfeldt-Jakob Disease, SB, salt bridge; HB, hydrogen bond(ed); saPMCA, serial automated protein misfolding cyclic amplification; PDB Bank, protein data bank (http://www.rcsb.org/); H1 or α1, α-Helix 1; H2 or α2, α-Helix2; H3 or α3, α-Helix 3; β1, β-strand 1; β2, β-strand 2; MD, Molecular Dynamics.

# Introduction

It has been a challenge to rational whether the contagious Transmissible Spongiform Encephalopathy (TSE) is caused by prions (Prusiner, 1997 & 1998; Soto & Castilla, 2004; Soto, 2011; Fernandez-Borges et al., 2012). As a misfolded protein, prion is neither a virus, nor a bacterium, and nor any microorganism. Prion disease cannot be caused by the vigilance of the organism immune system but it can freely spread from one species to another species. Humans TSEs (for example, Creutzfeldt-Jakob Disease (CJD) and variant CJD (vCJD)) can happen randomly through a number of processes, such as infections of transplanted tissue, blood transfusions and/or consumption of infected beef products, etc.. Many mammals such as cat, mink, deer, elk, moose, sheep, goat, nyala, oryx, greater kudu and ostrich etc. are also susceptible to TSEs. However, a small group of other animals such as rabbits, horses and dogs seem to be little affected by prions (Vorberg, Martin, Eberhard, & Suzette, 2003; Khan et al., 2010; Polymenidou et al., 2008; Zhang, 2011a; Zhang & Liu, 2011). As a result, it is important to understand and to identify the specific causes why these animals are unlikely to be affected by prions, as it will provide insight to prion diseases and help to resolve the prion diseases issue.

The role of $PrP^{Sc}$ infection in animals such as rabbit has been subject to a heated debating. A number of previous studies showed that a few animals such as rabbits exhibit low susceptibility to be infected by the $PrP^{Sc}$ (Vorberg et al., 2003; Khan et al., 2010; Barlow & Rennie, 1976; Fernandez-Funez et al., 2009; Korth et al., 1997; Courageot et al., 2008; Vilette et al., 2001; Nisbet et al., 2010; Wen et al., 2010a & 2010b; Zhou et al., 2011; Ma et al., 2012).

Now experimental structural data for rabbit $PrP^C$ ($RaPrP^C$) is available from the Protein Data Bank (PDB bank: http://www.rcsb.org/). For example, the structures of $RaPrP^C$ obtained from NMR (Li et al., 2007, with a PBD entry of 2FJ3) and X-ray (Khan et al., 2010, with a PBD entry of 3O79) measurements. As a result, it is desirable to reveal the properties and specific mechanisms of the

RaPrP$^C$ and the conversion process of PrP$^C$→PrP$^{Sc}$ of rabbits from limited experimental results. Here PrP$^C$ is a soluble normal cellular prion protein and PrP$^{Sc}$ is insoluble abnormally folded infectious and diseased prions. The present study will base on the X-ray and NMR structure of RaPrP$^C$ using molecular dynamics (MD) simulation techniques. The information from the present MD studies is able to provide valuable insight for the PrP$^C$→PrP$^{Sc}$ conversion. The information will provide useful rational in the design of novel therapeutic approaches and drugs that stop the conversion and disease propagation.

This paper can capture the α-helices→β-sheets conversion of PrP$^C$→PrP$^{Sc}$ under pH environments from neutral to low. The removing of salt bridges (SBs) under low pH environment can lead to this conversion. The 2FJ3.pdb of the RaPrP$^C$ has the distance of ASP177-ARG163 (O-N) about 10 Å which is nearly null in terms of energy. Thus, this SB does not exist and will not at all contribute to the PrP$^C$→PrP$^{Sc}$ conversion. This SB just links the β2-α2 loop, which has been a focus in PrP studies (Sweeting et al., 2009 & 2013; Wen et al., 2010b; Christen et al., 2008 & 2009 & 2012 & 2013; Damberger et al., 2011; Sigurdson et al., 2009 & 2010 & 2011; Pérez et al., 2010; Gossert et al., 2005; Lührs et al., 2003; Stankeret al., 2012; Cong et al., 2013; Bett et al., 2012; Meli et al., 2011; Rossetti et al., 2010 & 2011; Kirby et al., 2010; Zhang, 2011b & 2012b; Gorfe & Caflisch, 2007; Tabrett et al., 2010). Hence, our SB has caused much more debate, and it is very necessary to us to specially organize a paper to address this problem.

The rest of this paper is organized as follows. The MD simulation materials and MD reproducible methods for the X-ray and NMR structures of RaPrP$^C$ wild-type are provided in next section, followed by the analysis and discussion focusing on the MD trajectory results of the SB between ASP177 and ARG163 linking the β2-α2 loop and their discussions. Numerous new MD results of X-ray RaPrP$^C$ structure will be reported in the Results and Discussion section. Finally, a concluding remark on this SB, the β2-α2 loop and the interaction of this loop with the tail of H3 of PrP$^C$ is summarized.

## Materials and methods

The materials, e.g., data used in the present study are based on the laboratory NMR and X-ray PDB files of 2FJ3.pdb, 3O79.pdb respectively. For the NMR structure RaPrP$^C$(124-228), the experimental temperature is 298 K, pH value is 4.5, and pressure is 1 ATM; for the X-ray structure RaPrP$^C$(126-230), the resolution is 1.60 Å, R-value is 0.161, R-free is 0.218, temperature is 100 K, pH value is 6.5 (where, possibly owing to the different pH conditions, in X-ray structure 3O79.pdb we can observe the SB Asp177-Arg163 (O-N) but in NMR structure 2FJ3.pdb we cannot).

The MD methods employed are the same as the previous studies (Zhang & Zhang, 2013 & 2014; Zhang, 2010 & 2011c). Briefly, all simulations used the ff03 force field of the AMBER 11 package (Case et al., 2010). The systems were surrounded with a 12 Å layer of TIP3PBOX water molecules and neutralized by sodium ions using the XLEaP module of AMBER 9. To remove the unwanted bad contacts, the systems of the solvated proteins with their counter ions had been minimized mainly by the steepest descent method and followed by a small number of conjugate gradient steps on the data, until without any amino acid clash checked by the Swiss-Pdb Viewer 4.1.0 (http://spdbv.vital-it.ch/). Next, the solvated proteins were heated from 100 K to 300 K in 1 ns duration. Three sets of initial velocities denoted as seed1, seed2 and seed3 are performed in parallel for stability (this will make each set of MD starting from different MD initial velocity, implemented in Amber package we choose three different odd-real-number values for "ig") – but for the NMR structure and the X-ray structure of RaPrP$^C$, each set of the three has the same "ig" value in order to be able to make comparisons. The thermostat algorithm used is the Langevin thermostat algorithm in constant NVT ensembles. The SHAKE algorithm (only on bonds involving hydrogen) and PMEMD (Particle Mesh Ewald Molecular Dynamics) algorithm with non-bonded cutoff of 12 Å were used during heating. Equilibrations were reached in constant NPT ensembles under Langevin thermostat for 5 ns. After equilibrations, production MD phase was carried out at 300 K for 30 ns using constant pressure and temperature ensemble and the PMEMD algorithm with the same non-bonded cutoff of 12 Å during simulations. The step size for equilibration was 1 fs and 2 fs in the MD production runs. The structures were saved to file every 1000 steps. During the constant NVT and then NPT ensembles of PMEMD, periodic boundary conditions have been applied.

In order to obtain the low pH (acidic) environment, the residues HIS, ASP, GLU were changed into their zwitterion forms of HIP, ASH, GLH, respectively, and Cl- ions were added by the XLEaP module of the AMBER package. Thus, the SBs of the system (residues HIS, ASP, GLU) under the neutral pH environment were broken in the low pH environment (zwitterion forms of HIP, ASH, GLH).

# Results and discussion

### Confirmation from the X-ray structure 3O79.pdb file

The 3O79.pdb file of RaPrP$^C$ X-ray structure consists of A and B chains. We can confirm as follows there is the SB ASP178-ARG164 (O-N) in both the A chain and B chain molecule structures. In A-chain structure, we find the SBs illuminated in the 1st graph of Fig. 1 and listed in Tab. 1, where the cut-off distance for a SB is 4 Å (Kumar & Nussinov, 2002) according to the definition of a SB.

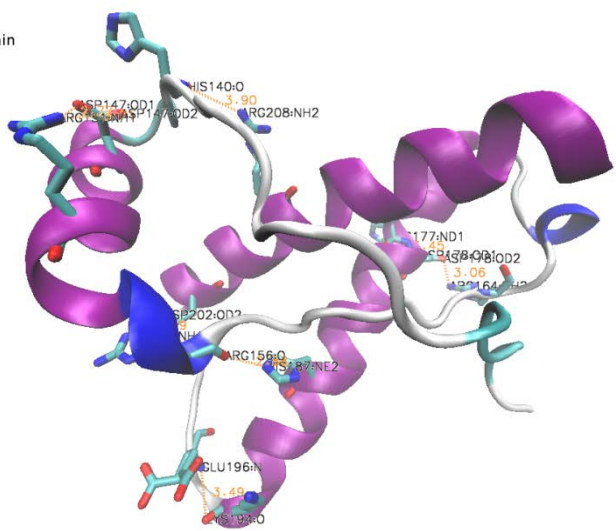

A-chain

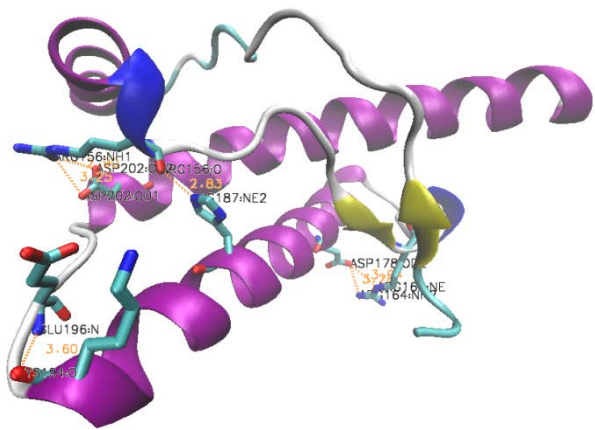

B-chain

**Fig. 1**: All the SBs in the A-chain structure and B-chain structure of the X-ray PDB file of RaPrP$^C$ (3O79.pdb). The orange dashed lines denote the SBs and the yellow data denote the distances of the SBs.

**Tab. 1**: All the SBs in the A-chain structure of the X-ray PDB file of RaPrP$^C$ (3O79.pdb)

| O atoms | N atoms | Distance (Å) |
|---|---|---|
| ASP178.OD2 | ARG164.NH2 | 3.06 Å |
| ASP202.OD2 | ARG156.NH1 | 2.89 Å |
| ARG156.O | HIS187.NE2 | 2.88 Å |
| LYS194.O | GLU196.N | 3.49 Å |
| ASP178.OD1 | HIS177.ND1 | 3.45 Å |
| HIS140.O | ARG208.NH2 | 3.90 Å |
| ASP147.OD1 | ARG151.NH1 | 2.77 Å |
| ASP147.OD2 | ARG151.NH1 | 3.71 Å |

**Tab. 2**: All the SBs in the B-chain structure of the X-ray PDB file of RaPrP$^C$ (3O79.pdb)

| O atoms | N atoms | Distance (Å) |
|---|---|---|
| <u>ASP178.OD2</u> | <u>ARG164.NE</u> | <u>3.94 Å</u> |
| <u>ASP178.OD2</u> | <u>ARG164.NH2</u> | <u>3.28 Å</u> |
| <u>ASP202.OD1</u> | <u>ARG156.NH1</u> | <u>3.25 Å</u> |
| <u>ASP202.OD2</u> | <u>ARG156.NH1</u> | <u>2.90 Å</u> |
| ARG156.O | HIS187.NE2 | 2.83 Å |
| LYS194.O | GLU196.N | 3.60 Å |

In B-chain structure, we find the SBs illuminated in the 2nd graph of Fig. 1 and listed in Tab. 2. Our MD experiences showed that the underlined (in real and dash lines) SBs are two basic and fundamental SBs of RaPrP$^C$.

## Confirmation from the MD of X-ray 3O79.pdb structure at 300 K room temperature

We denote the three α-helices and the two β-strands of PrP$^C$ as β1, α1, β2, α2 and α3. The MD results from the simulations are summarized in Fig.s 2-6, under the parallel conditions of seed1, seed2 and seed3 in the duration of 30 ns.

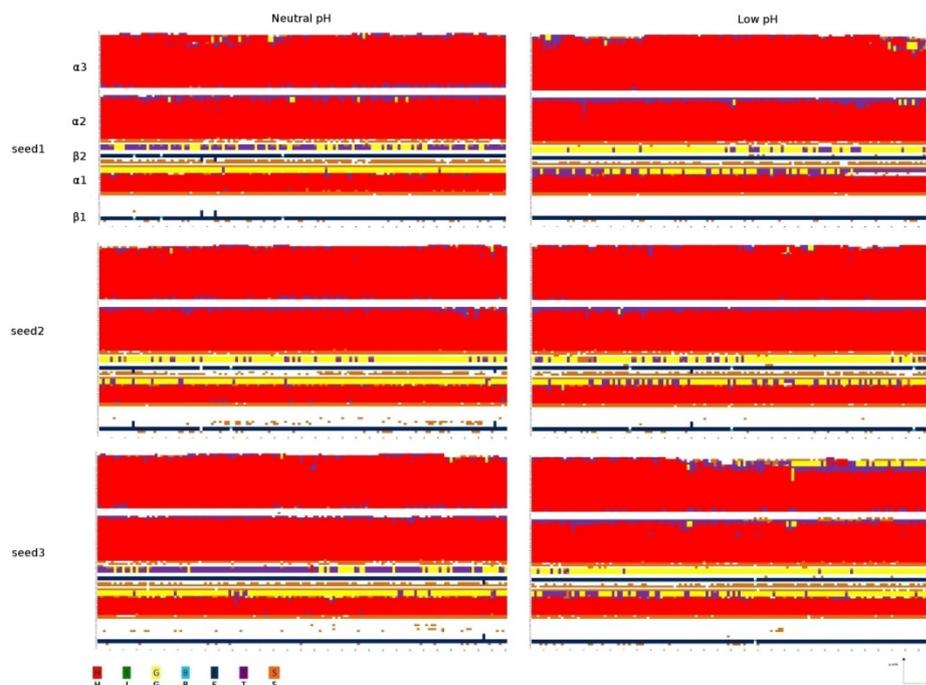

**Fig. 2:** Secondary Structure graphs for X-ray RaPrP$^C$ wild-type at 300 K (x-axis: time (0-30 ns), y-axis: residue number (126–230); left column: neutral pH, right column: low pH; up to down: seed1-seed3. H is the α-helix, I is the π-helix, G is the 3-helix or $3_{10}$ helix, B is the residue in isolated β-bridge, E is the extended strand (participates in β-ladder), T is the hydrogen bonded turn, and S is the bend.

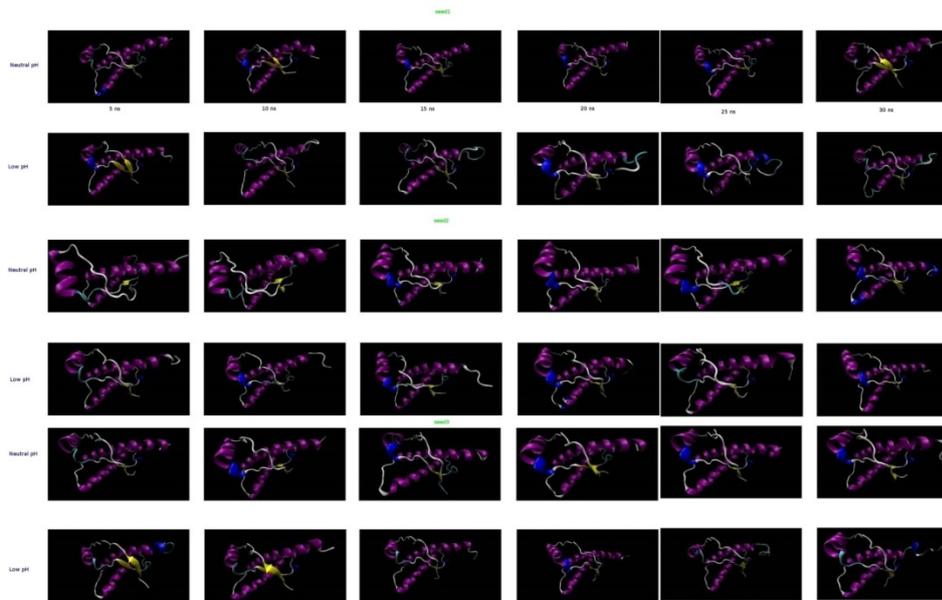

**Fig. 3:** Snapshots at 5 ns, 10 ns, 15 ns, 20 ns, 25 ns and 30 ns (from left to right in turns) for MD of X-ray structure of RaPrP$^C$ wild-type at 300 K. The 1$^{st}$ and 2$^{nd}$ rows are for seed1, the 3$^{rd}$ and 4$^{th}$ rows are for seed2, and the 5$^{th}$ and 6$^{th}$ rows are for seed3. The 1$^{st}$, 3$^{rd}$ and 5$^{th}$ rows are for neutral pH environment, and the 2$^{nd}$, 4$^{th}$, and 6$^{th}$ rows are for low pH environment.

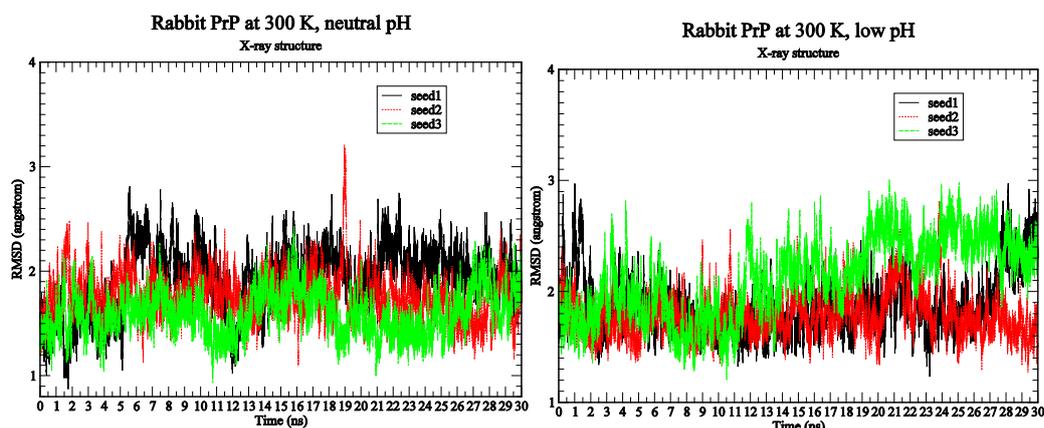

**Fig. 4:** RMSD (root-mean-square deviation) graphs for X-ray RaPrP$^C$ wild-type at 300 K (x-axis: time (0-30 ns), y-axis: RMSD value (angstrom); left column: neutral pH, right column: low pH; seed1: black real line, seed2: red dot line, seed3: green dashed line).

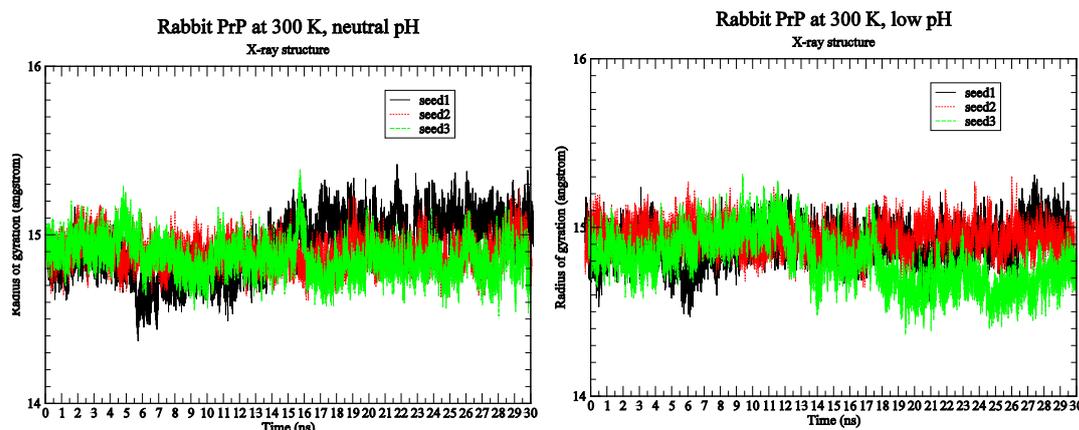

**Fig. 5:** Radius of Gyration graphs for X-ray RaPrP$^C$ wild-type at 300 K (x-axis: time (0-30 ns), y-axis: Radius of gyration value (angstrom); left column: neutral pH, right column: low pH; seed1: black real line, seed2: red dot line, seed3: green dashed line).

From Fig.s 2 & 3, we cannot see gigantic changes of secondary structures under pH environments from neutral to low; but we found the following changes: (i) the 'tails' (i.e. C-terminus) of α2 and α3 have α-helix unfolded into HB turns, bends and $3_{10}$-hleices, for seed2 a little part of α2 unfolded into β-bridge, (ii) some parts of the yellow coloured $3_{10}$-hleices between α1 and β2 become into purple HB turns, (iii) for seed1 and seed3 some parts of the purple coloured HB turns between β2 and α2 become into yellow coloured $3_{10}$-hleices, (iv) for seed1 the head of α2 becomes into bends, (v) for some snapshots their two β-strands β1 & β2 (i.e. the β-sheet) become longer. Fig. 3 shows the snapshots of Fig. 2 at 5 ns, 10 ns, 15 ns, 20 ns, 25 ns and 30 ns. In Fig. 3, we can see there is a short $3_{10}$-helix in the β2-α2 loop under both neutral and low environments, but for seed1 the S174N mutation (4HMM.pdb) made this $3_{10}$-helix disappear at 30 ns; during the long 30 ns of MD simulations in neutral pH environment, for both the wild-type and the S174N mutant, we found in the β2-α2 loop there are 3 hydrophobic contacts/cores Pro165-Val166-Phe175-Val176 with occupied rate greater than 95%, and 2 hydrophobic contacts Pro165-Met129, Val166-Ala225 with high occupied rate. Being different from wild-type, the S174N mutation produced two new hydrophobic contacts Val176-Val180, Val176-Ile215 with 100% occupied rate during the whole 30 ns of seed1~seed3 MD simulations. In the β2-α2 loop, for the wild-type at 300 K under neutral pH environment, there always exist 3 HBs Asp178-Arg164 (see Tab. 3 for the occupied rates), Cys179-Thr183 (with occupied rates 82.83%, 72.79%, 84.29% for seed1~seed3 respectively), Asp178-Tyr169 (with occupied rates 27.44%, 33.93%, 5.09% for seed1~seed3 respectively); there also exist some special HBs Gln172-Gln219 (with occupied rates 8.07%, 7.39% for seed1, seed3 respectively), Val166-Ser170 (with occupied rate 7.01% for seed3), Tyr169-Tyr218 (with occupied rate 5.51% for seed3), Tyr218-Ser170 (with occupied rate 48.13% for seed3), Ser170-Asn171 (with occupied rate 16.76% for seed2), and Asn171-Ser174 (with occupied rate 23.92% for seed2). We found the S174N mutation made the loss of the strong HB Cys179-Thr183. Other HBs such as Asp202-Tyr149, Asp202-Arg156, Asp202-Tyr157, and Lys204-Glu146 are keeping H3 and H1 linked; at the same time, H1 linking with H2 and the H2-H3 loop are kept by two strong SBs Arg156-His187 and Arg156-Glu196. For SBs, we found the S174N mutation has not greatly changed SBs of the wild-type; the occupied rates of SB Asp178-Arg164 for the wild-type are 70.59%, 26.27%, 61.89% for seed1~seed3 respectively.

**Tab. 3**: The occupied rates for HB Asp178-Arg164 of the wild-type (X-ray structure) at 300 K, neutral pH value during the 30 ns of MD simulations.

| HB | seed1 | seed2 | seed3 |
|---|---|---|---|
| ASP178@OD1-ARG164@NH2.HH21 | 51.01% | 52.47% | 59.76% |
| ASP178@OD2-ARG164@NH2.HH21 | 44.65% | 53.73% | 38.06% |
| ASP178@OD1-ARG164@NE.HE | 30.57% | 13.92% | 10.00% |
| ASP178@OD2-ARG164@NE.HE | 31.73% | 14.91% | 45.49% |

The above analysis means that for the wild-type X-ray RaPrP structures we can capture the α-helices→β-sheets conversion of $PrP^C$→$PrP^{Sc}$ under pH environments from neutral to low, because of the removing of SB networks of neutral pH environment. The SB ASP178-ARG164 (O-N) (Fig. 6) is one of these important SBs and seeing Fig. 6 we know it always exists during 30 ns. For seed1~seed3, SBs ASP178.OD1-ARG164.NE and ASP178.OD2-ARG164.NE are always existing during the whole 30 ns, SBs ASP178.OD1-ARG164.NH2 and ASP178.OD2-ARG164.NH2 are always strongly existing during the whole 30 ns, but SBs ASP178.OD1-ARG164.NH1 and ASP178.OD2-ARG164.NH1 are not existing at all during the whole 30 ns.

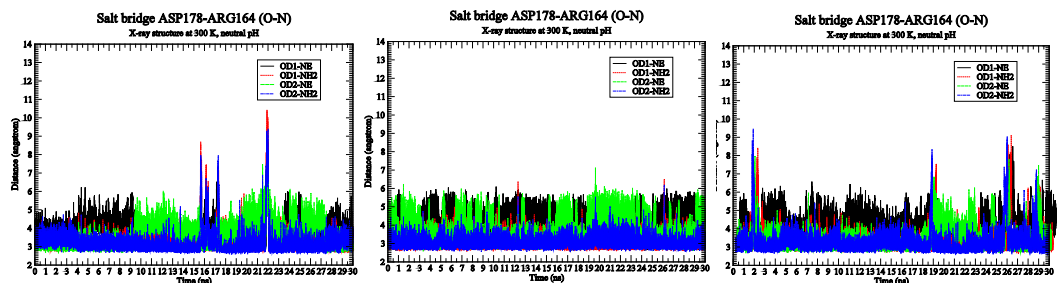

**Fig. 6:** The occupying of SB ASP178-ARG164 (O-N) during the 30ns of MD simulations of the wild-type X-ray structure RaPrP at 300 K under neutral pH environment for seed1-seed3 [seed1-seed3: from left to right in turns].

If we set the distance cut-off 3.0 Å and the angle cut-off 20 degrees for the hydrogen bonds (HBs) as in the VMD package (http://www.ks.uiuc.edu/Research/vmd/), seeing the last row of Fig. 4 in (Zhang & Zhang, 2014), we know that the HB between ASP178 and ARG164 is almost 100% there under neutral pH environment, and it becomes very much weaker under low pH environment because of the removing of the electrostatic interactions of the SB between ASP178 and ARG164.

Regarding the RMSD (Fig. 4) and Radius Of Gyration (Fig. 5) of the MD simulations of X-ray structure, we can see from Fig.s 4~5 that the 30 ns simulations is short but good enough for a small protein like the prion protein. We also can see from Fig.s 4~5 that the three seeds are valid and there are not great differences among them.

## Confirmation from the MD of the NMR 2FJ3.pdb structure at 300 K room temperature

The secondary structure changes for the NMR wild-type RaPrP at room temperature 300 K can be seen in Fig. 2 of (Zhang & Zhang, 2014). As can be seen from Fig. 2 of (Zhang & Zhang, 2014), the left panels are dominated by red colour, indicating that under the neutral pH conditions, the three α-helices (α1, α2 and α3) of the wild-type prion RaPrP$^C$, remain dominant the α-helices without significant changes during the period of 30 ns, regardless the seed conditions. For example, the α-helices (α1, α2 and α3 in red colour) of the top-left panel does not experience any apparent colour changes, indicating that under the neutral pH condition, the α-helices of the wild-type prion resist structural changes. This is particular the case in α3 and α2, although small noticeable changes in α1 has been observed. However, on the right hand side of the figure, under acidic condition (i.e. low pH environment) the seed1-seed3cases of the wild-type prion protein show changes indicated by their changes of colour codes. In low pH environment, the wild-type prion protein turns into a colourful panel but (i) large presentation of HB bends for α3 in seed1-seed3 and α2 in seed2, (ii) $3_{10}$-helix for α2 in seed3 and α1 in seed1-seed3, and (iii) β1 and β2 become longer under acidic condition for seed2 and seed3, for seed1 the β2 becomes into β-bridge structures. Therefore, under acidic condition (low pH environment), the SB network of the wild-type (RaPrP$^C$) is broken thus leads to the unfolding of the stable α-helical structures of RaPrP$^C$. Hence, it suggests that the structural distributions of the wild-type (RaPrP$^C$) protein depends on the pH. Generally, we could see the clear secondary structure changes from the unfolding of the α-helical structures conversion into β-sheets from Fig. 2 of (Zhang & Zhang, 2014). If we set the HB distance cut-off be 3.0 Å, seeing the first row of Fig. 4 of (Zhang & Zhang, 2014), we know that the HB between ASP177 and ARG163 does not exist for seed3, exists occasionally for seed2, and will not exist after 19 ns for seed1. However, the following Fig. 5 can still confirms there is a SB ASP177-ARG163 (O-N).

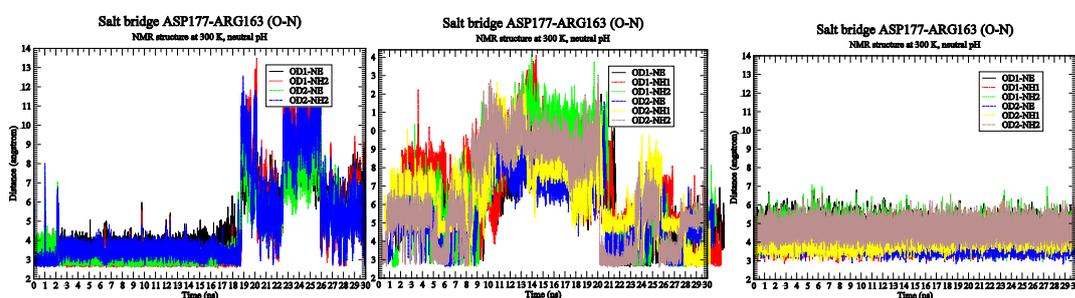

**Fig. 7:** The occupying of SB ASP177-ARG163 (O-N) during the 30 ns of MD simulations of the wild-type NMR structure RaPrP at 300 K under neutral pH environment for seed1-seed3 [seed1-seed3: from left to right in turns].

By Fig. 7, (i) for seed1 (Fig. 7 graph 1), the SBs ASP177.OD1-ARG163.NE and ASP177.OD1-ARG163.NH2 exist until 18.742 ns, the SB ASP177.OD2-ARG163.NE exists until 18.724 ns and the SB ASP177.OD2-ARG163.NH2 exists until 18.744 ns; (ii) for seed2 (Fig. 7 graph 2), generally speaking, except for during 9~20 ns the SBs of ASP177-ARG163 (O-N) exist from 0 ns to 30 ns, SB ASP177.OD1-ARG163.NE exists during 0~8.03 ns and 20.108~30 ns (where the gap might be due to the SBs ASP177-HIS176 and ASP166-ARG163 drawing ASP177 and ARG163 away respectively), SB ASP177.OD1-ARG163.NH1 exists during 7.678~8.032 ns and 27.572~30 ns and occasionally during 23.38~23.88 ns, SB ASP177.OD1-ARG163.NH2 exists during 0~8.052 ns and 20.11~27.628 ns, SB ASP177.OD2-ARG163.NE exists during 0~8.168 ns and 20.108~29.986 ns, SB ASP177.OD2-ARG163.NH1 exists during 7.672~7.908 ns and 27.572~29.992 ns, and SB ASP177.OD2-ARG163.NH2 exists during 0~8.054 ns and 20.11~27.69 ns; and (iii) for seed3 (Fig. 7 graph 3), weak SB ASP177.OD1-ARG163.NE always exists during 0~30 ns, very strong SB ASP177.OD1-ARG163.NH1 always exists during 0~30 ns, strong SB ASP177.OD1-ARG163.NH2 always exists during 0~30 ns, very strong SB ASP177.OD2-ARG163.NE always exists during 0~30 ns, very strong SB ASP177.OD2-ARG163.NH1 always exists during 0~30 ns, and strong SB ASP177.OD2-ARG163.NH2 always exists during 0~30 ns.

From the NMR structure 2FJ3.pdb file, RaPrP$^C$ has no SB ASP177-ARG163 (O-N) because of the following large distances (≈10.2 Å) of ASP177-ARG163 (O-N) (Tab. 4): 10.73 Å for ASP177.OD1-ARG163.NE, 12.13 Å for ASP177.OD1-ARG163.NH1, 10.63 Å for ASP177.OD1-ARG163.NH2, 8.70 Å for ASP177.OD2-ARG163.NE, 10.26 Å for ASP177.OD2-ARG163.NH1, and 8.85 Å for ASP177.OD2-ARG163.NH2. For the 2FJ3.pdb, we found other SBs of RaPrP$^C$ (Tab. 5): ASP201.OD1-ARG155.NH1 with distance 2.87 Å, ASP201.OD1-ARG155.NH2 with distance 2.55 Å, ASP146.OD2-LYS139.O with distance 2.82 Å, ASP143.O-GLU145.N with distance 3.76 Å, GLU151.OE2-ARG155.NH1 with distance 3.67 Å, GLU151.OE2-ARG155.NE with distance 3.25 Å, GLU195.OE2-LYS193.NZ with distance 2.93 Å, ARG155.O-LYS186.O with distance 3.06 Å, ARG155.O-LYS186.ND1 with distance 2.79 Å, GLU210.OE1-LYS176.ND1 with distance 3.97 Å, and GLU210.OE2-LYS176.ND1 with distance 2.60 Å.

**Tab. 4**: The O-N distances of ASP177-ARG163 of the NMR PDB file of RaPrP$^C$ (2FJ3.pdb)

| O atoms | N atoms | Distance (Å) |
|---|---|---|
| ASP177.OD1 | ARG163.NE | 10.73 Å |
| ASP177.OD1 | ARG163.NH1 | 12.13 Å |
| ASP177.OD1 | ARG163.NH2 | 10.63 Å |
| ASP177.OD2 | ARG163.NE | 8.70 Å |
| ASP177.OD2 | ARG163.NH1 | 10.26 Å |
| ASP177.OD2 | ARG163.NH2 | 8.85 Å |

**Tab. 5**: All the SBs in the NMR structure of the PDB file of RaPrP$^C$ (2FJ3.pdb)

| O atoms | N atoms | Distance (Å) |
|---|---|---|
| ASP201.OD1 | ARG155.NH1 | 2.87 Å |
| ASP201.OD1 | ARG155.NH2 | 2.55 Å |
| ASP146.OD2 | LYS139.O | 2.82 Å |
| ASP143.O | GLU145.N | 3.76 Å |
| GLU151.OE2 | ARG155.NH1 | 3.67 Å |
| GLU151.OE2 | ARG155.NE | 3.25 Å |
| GLU195.OE2 | LYS193.NZ | 2.93 Å |
| ARG155.O | LYS186.O | 3.06 Å |
| ARG155.O | LYS186.ND1 | 2.79 Å |
| GLU210.OE1 | LYS176.ND1 | 3.97 Å |
| GLU210.OE2 | LYS176.ND1 | 2.60 Å |

**Confirmation from the MD of the NMR 2FJ3.pdb structure at 350 K and 450 K**

Seeing Tab. 1 of (Zhang & Zhang, 2014), for the SB ASP177-ARG163 (O-N), we know its occupied rates for seed1-seed3 are 47.80%, 40.38%, 21.92% at 450 K, and 19.54%, 6.09%, 38.69% at 350 K. We observed from the second and third rows of Fig. 4 of (Zhang & Zhang, 2014) that ASP177.OD1/2-ARG163.NE/NH2 should exist at 350 K and 450 K under neutral pH environment for RaPrP$^C$. In conclusion, by Tab. 1 and Fig. 4 of (Zhang & Zhang, 2014) the SB ASP177-ARG163 (O-N) exists at 350 K & 450 K for RaPrP$^C$.

In Fig. 3 of (Zhang & Zhang, 2014), we can see that percentages of β-sheet are clearly increasing under low pH environment at 350 K for seed1 and seed3 (for seed1 the loop between α2 and α3 becomes into β-ladders and for seed3 the β2-α2 loop becomes into β-ladders and β-bridges), and the three α-helices are unfolded into a colourful panel of almost all of the seven colours but large presentation of yellow 310-helices, purple HB turns and orange bends for seed1-seed3. In Fig. 1 of (Zhang & Zhang, 2014), under low pH environment for seed2 the two β-strands of β-sheet clearly become longer, for seed1-seed3 some parts of the three α-helices become into β-sheet structures, and the three α-helices are unfolded into a colourful panel of almost all of the seven colours but large presentation of orange bends for seed1-seed3. This reason to cause the above α-helices→β-sheets conversion is due to the removing of SBs (by changing residues HIS, ASP, GLU into their zwitterion forms of HIP, ASH, GLH in order to obtain the low pH (acidic) environment). SB ASP177-ARG163 (O-N) is one of the SBs.

### Confirmation from the MD of the homology structure of RaPrP$^C$ at 500 K

For RaPrP$^C$, its X-ray structure (3O79.pdb) was released into PDB bank on 2010-11-24 and its NMR structure (2FJ3.pdb) was released into PDB bank on 2006-12-31. Earlier in 2004, CSIRO scientists built a homology model for RaPrP$^C$ (denoted as 6EPA.pdb). The homology structure is for RaPrP$^C$(120-229) made by mutations using the NMR structure of human PrP$^C$(125-228) (1QLX.pdb) (Zhang et al., 2006). We did MD for the homology structure at 500 K and found that SB ASP177-ARG163 (O-N) is conserved through a large part of the 30 ns' simulations (Zhang et al., 2006).

Interestingly, the SB ASP177-ARG163 (O-N) does not affect the structures of human and mouse prion proteins very much (Zhang, 2011b). Kuwata et al. (2007) presented an anti-prion drug GN8 fixing the distance between N159 and E196 being 1.54 Å. Thus, we might propose to fix the distance between ASP177 and ARG163 in a SB distance to design an anti-prion drug (this proposal might be wrong but this SB at β2-α2 loop plays a very important role for the stability of RaPrP$^C$).

## Conclusion

This paper mainly discusses the presence/influence of a SB in RaPrP in the context of a possible role in PrP toxicity. We compare the X-ray crystal structure and NMR structure. Of note, both structures were obtained under different pH conditions, which affect electrostatics and structure. It has to be considered that presence of a salt bridge depends on the environmental condition. Hence, the topic itself is interesting us and some furthermore study directions (i)-(iii) are pointed to us. (i). Given the dependence of SBs on the environment, the statement made on some scientists claiming a low energy contribution for a 10 Å SB at pH 4.5 thereby challenging us results might be wrong is not valid if justified by a structure obtained at different pH environment – MD studies of many and dynamic structures will be a way to confirm some observations. (ii). Because NMR and X-ray structures were got under different environments, in order to make comparisons we put them into the same environment to make MD calculations, this might be scientific in theory but not completely agreeing with the X-ray environment – we will furthermore do MD studies along the way of the X-ray environment. To get a clue on the protonation states at the different pH values the electrostatics should be calculated for the starting structure, as geometry may result in rather strong pKa deviations of individual groups. (iii). For the X-ray structure and NMR structure, different starting geometries sampled or different starting velocities should be furthermore studied for the MD theory.

There was a big lot of controversy over "prion" theory and recently on saPMCA produced rabbit prions. We have found one focus of prion protein structures is at the β2-α2 loop and its interacted C-terminal of H3 (Biljan et al., 2012a & 2012b & 2011; Calzolai et al., 2000; Christen et al., 2009 & 2012; Damberger et al., 2011; Gossert et al., 2005; Ilc et al., 2010; Lee et al., 2010; Wen et al., 2010a & 2010b; Kong et al., 2013; Perez et al., 2010 & 2008; Sweeting et al., 2013; Zahn et al., 2003; Zhang et al., 2000; Kurt et al., 2014a & 2014b). This article found there is a SB ASP177-ARG163 (O-N) in RaPrP$^C$, which just keeps this loop being linked. The function of this SB linkage should be furthermore confirmed in experimental laboratories.

## Supplementary material

The supplementary material of movies for 30 ns' MD trajectories of NMR and X-ray RaPrP$^C$ at room temperature 300 K with the SB ASP177-ARG163 (O–N) will be available online at:

<p align="center">https://sites.google.com/site/jiapuzhang/</p>

## Acknowledgments

This research has been supported by a Victorian Life Sciences Computation Initiative (VLSCI) grant numbered VR0063 on its Peak Computing Facility at the University of Melbourne, an initiative of the Victorian Government of Australia.